\documentclass[preprint,letter]{jpsj2}

\title{$^1$H-NMR Spin-Lattice Relaxation Rate of the Quantum Spin System 
(CH$_3$)$_2$CHNH$_3$Cu(Cl$_{x}$Br$_{1-x}$)$_3$ with $x=0$ and $0.35$.}

\author{Akira \textsc{Oosawa}, Takao \textsc{Suzuki}$^1$,
Keishi \textsc{Kanada}, Satoshi \textsc{Kojima}, Takayuki \textsc{Goto} 
and Hirotaka \textsc{Manaka}$^2$}

\inst{Department of Physics, Sophia University, 7-1 Kioi-cho, Chiyoda-ku, Tokyo 102-8554, Japan\\
$^1$Advanced Meson Science Laboratory, RIKEN Nishina Center, 2-1 Hirosawa, 
Wako, Saitama 351-0198, Japan \\
$^2$Graduate School of Science and Engineering, Kagoshima University, Korimoto, 
Kagoshima 890-0065, Japan\\}

\recdate{\today}

\abst{
The spin-lattice relaxation rate $T_1^{-1}$ of $^1$H-NMR has been measured in 
(CH$_3$)$_2$CHNH$_3$Cu(Cl$_x$Br$_{1-x}$)$_3$ with $x=0$ and $0.35$, 
in order to investigate their microscopic magnetism. 
Previous macroscopic magnetization and specific heat measurements
suggested that these two exist in a singlet-dimer phase. 
The temperature dependence of $T_1^{-1}$ in an $x=0$ system decreased exponentially toward zero, 
providing microscopic evidence of a gapped singlet ground state,
which is consistent with results of macroscopic experiments. 
At the same time, in the $x=0.35$ system, $T_1^{-1}$ showed 
a sharp peak structure at approximately 7.5 K 
but no splitting of $^1$H-NMR spectra, 
indicative of magnetic ordering. 
We discuss the observed sharp peak structure in the $x=0.35$ system 
with the soft mode toward the exotic magnetic ground state suggested by 
recent $\mu$SR experiments.}

\kword{(CH$_3$)$_2$CHNH$_3$Cu(Cl$_x$Br$_{1-x}$)$_3$, quantum spin system, spin gap, 
bond randomness, $^1$H-NMR, spin-lattice relaxation rate $T_1^{-1}$}

\begin{document}
\sloppy
\maketitle

When a strong magnetic field or pressure is applied in spin gap systems, 
the gapped singlet ground state becomes magnetic owing to the vanishing of the spin gap 
so that the system can undergo magnetic ordering with the help of 
three-dimensional magnetic interactions. 
This ordering is the magnetic quantum phase transition from the spin gap phase to 
the magnetic ordered phase. 
The field-induced and pressure-induced magnetic orderings have been given much attention
from the viewpoint of the magnon Bose-Einstein condensation \cite{Nikuni} 
and the longitudinal magnetic excitation mode \cite{Matsumoto,Matsumoto2,Matsumoto3}, 
respectively.
Meanwhile, when bond randomness or site randomness 
is introduced to a spin gap system, the gapped singlet ground state is disturbed 
so that magnetic moments are induced in a singlet sea. 
At low temperatures, the induced magnetic moments cause magnetic ordering 
in some cases as observed in 
(Cu$_{1-x}$Mg$_x$)GeO$_3$ \cite{MasudaCu}, 
Cu(Ge$_{1-x}$Si$_x$)O$_3$ \cite{Regnault}, 
Sr(Cu$_{1-x}$Zn$_x$)$_2$O$_3$ \cite{Azuma}, 
Pb(Ni$_{1-x}$Mg$_x$)$_2$V$_2$O$_8$ \cite{Uchiyama}, 
Tl(Cu$_{1-x}$Mg$_x$)Cl$_3$ \cite{Oosawa}, 
Bi- and Sr-substituted Pb$_2$V$_3$O$_9$, \cite{Waki} 
and (Sr$_{14-x}$Ca$_x$)Cu$_{24}$O$_{41}$ \cite{Nagata}, 
while they may be localized in a singlet sea in other cases, 
such as the magnon Bose glass phase, as discussed in 
(Tl$_{1-x}$K$_x$)CuCl$_3$ \cite{OosawaTlK,Shindo,Suzuki,Fujiwara}. 

The title compound (CH$_3$)$_2$CHNH$_3$Cu(Cl$_x$Br$_{1-x}$)$_3$ 
(abbreviated as IPACu(Cl$_x$Br$_{1-x}$)$_3$) 
is the mixed system of the two spin gap systems IPACuCl$_3$ and IPACuBr$_3$. 
The magnitude of the excitation gap $\Delta$ between the singlet ground state 
and triplet excited states in IPACuCl$_3$ was estimated to be 17.1 - 18.1 K 
by magnetic susceptibility measurements \cite{Manakasus}. 
From the viewpoint of the crystal structure of IPACuCl$_3$, 
the origin of the spin gap was expected to be the $S=\frac{1}{2}$ 
ferromagnetic-antiferromagnetic alternating chain along the $c$-axis \cite{Manakasus}. 
However, recently, it has been suggested that IPACuCl$_3$ should be characterized as 
a spin ladder along the $a$-axis with strongly coupled ferromagnetic rungs, namely, 
an antiferromagnetic chain with an effective $S=1$ ``composite Haldane chain'',
and the excitation gap was re-estimated as 13.6 K 
by neutron inelastic scattering experiments \cite{Masuda}. 
Meanwhile, IPACuBr$_3$ has been characterized as 
an $S=\frac{1}{2}$ antiferromagnetic-antiferromagnetic 
alternating chain with a singlet dimer phase and 
an excitation gap $\Delta= 98$ K \cite{ManakaBr}. 
However, IPACuBr$_3$ may also be recharacterized as the spin ladder system 
when the neutron inelastic scattering experiments are carried out, 
in the same manner as IPACuCl$_3$. 

%%trim=left, bottom, right,up
%%h=here, t=top, b=bottom, p=separate figure page
%%clip, scale, width=0.9\linewidth
\begin{figure}[t]
\begin{center}
\includegraphics[trim=0.2cm 12.0cm 0.1cm 2.0cm, clip, width=80mm]
{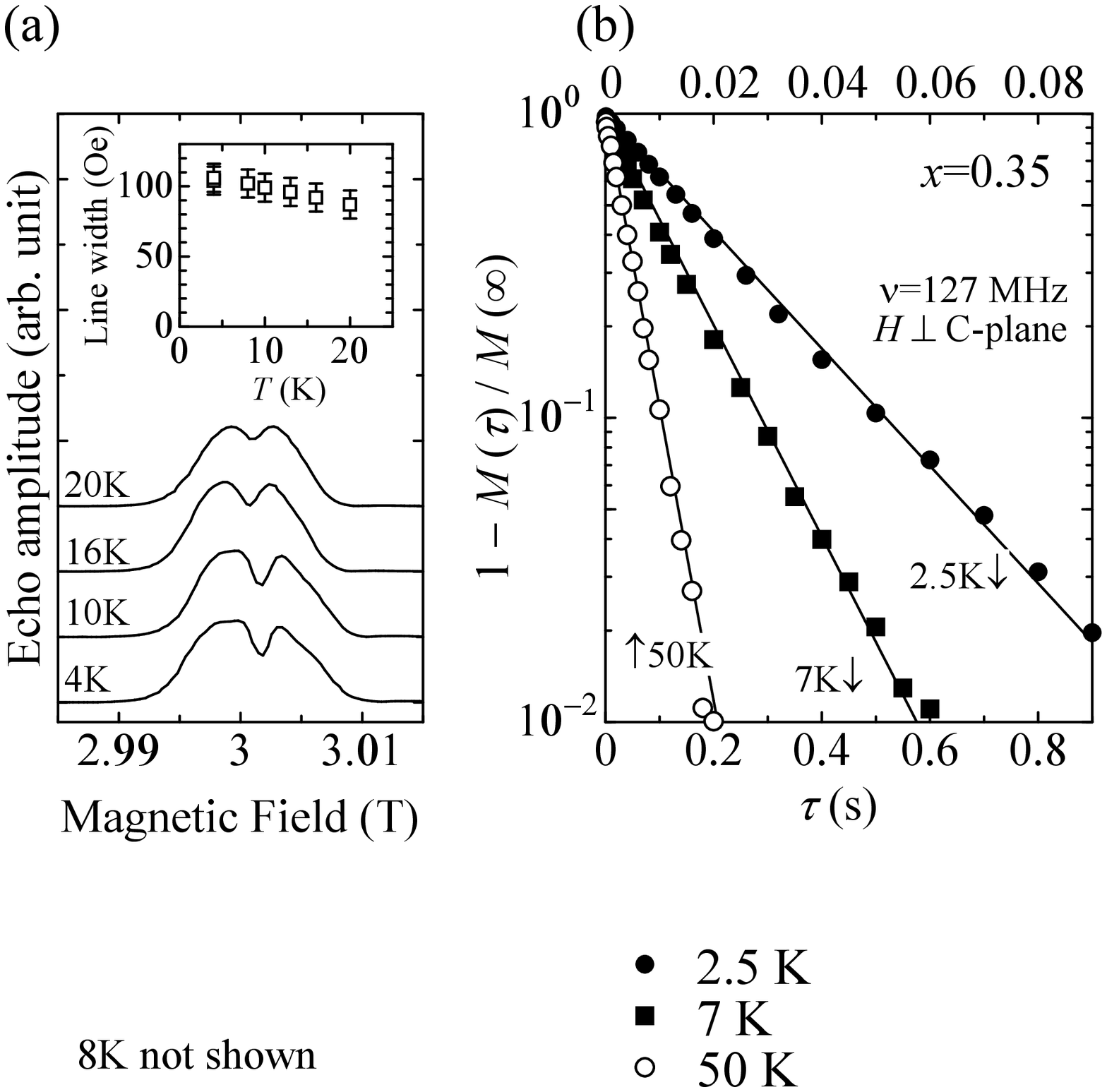}\\
%%{Fig0_2a.eps}\\
\end{center}
\caption{(a) Profile of $^{1}$H-NMR spectra at various temperatures
taken at $\nu=127$ MHz for $H \perp$ C-plane in IPACu(Cl$_{0.35}$Br$_{0.65}$)$_3$. 
The inset shows the temperature dependence of line width.
(b) Representative nuclear magnetization recovery of $^1$H-NMR at various temperatures. 
The solid lines denote the fits by Eq. (\ref{eq1}). \label{Fig1}}
\end{figure} 
The mixed system IPACu(Cl$_x$Br$_{1-x}$)$_3$ was studied 
by the magnetization and specific heat measurements, 
and it was reported that the bond-randomness-induced antiferromagnetic ordered phase 
with $T_{\rm N} =$ 13 - 17 K emerges in the region $0.44 < x < 0.87$ along 
with the Haldane phase in $x \ge$ 0.87 and the singlet-dimer phase 
in $x \le$ 0.44 \cite{Manakamixsus,Manakamixmag} 
and that the phase boundaries between these phases are of the first order. 
Recently, our group microscopically studied the mixed system 
IPACu(Cl$_x$Br$_{1-x}$)$_3$ by 
NMR \cite{Kanada,KanadaQuBS,Adachi} and $\mu$SR measurements\cite{Saito,Goto}. 
Microscopic evidence of the impurity-induced antiferromagnetic ordered phase, 
such as the clear splitting of the $^1$H-NMR spectra and rotation of the $\mu$SR time spectra below $T_{\rm N}$, 
was observed in IPACu(Cl$_x$Br$_{1-x}$)$_3$ with $x=0.85$. 
Furthermore, for a microscopic investigation on
the bond randomness effect in the Haldane phase, 
we measured the muon spin relaxation in IPACu(Cl$_x$Br$_{1-x}$)$_3$ with $x=0.95$ and 
a $^1$H-NMR spin-lattice relaxation rate $T_1^{-1}$ in IPACu(Cl$_x$Br$_{1-x}$)$_3$ with $x=0.88$. 
From these experiments, 
we concluded that the ground state becomes magnetic due to the bond randomness effect 
in the mixed system IPACu(Cl$_x$Br$_{1-x}$)$_3$ for $0.88 \le x <1$.  
This contrasts with the conclusion drawn from 
macroscopic experiments \cite{Manakamixsus,Manakamixmag}. 

In this Letter, we report and discuss the results of the $^1$H-NMR spin-lattice relaxation rate 
$T_1^{-1}$ in IPACu(Cl$_x$Br$_{1-x}$)$_3$ with $x=$ 0 and 0.35 
in order to microscopically investigate the bond randomness effect 
in the Br-rich region, 
which was reported to be of the singlet dimer phase. 

Single crystals of IPACu(Cl$_x$Br$_{1-x}$)$_3$ with $x=0$ and $0.35$ were prepared by 
slow evaporation\cite{Manakamixsus}. 
Crystals with three orthogonal surfaces were obtained. 
These three planes were termed A-, B-, and C-planes\cite{Manakasus,Torque}. 
Crystals were typically approximately $2 \times 3 \times 8$ mm$^3$.

The spin-lattice relaxation rate $T_1^{-1}$ of $^1$H-NMR was measured by 
saturation recovery with a pulse train.
There are ten inequivalent proton sites in the unit cell of the present system. 
Each inequivalent proton site has a different distance 
such that it is exposed to different hyperfine fields from the nearest magnetic Cu site. 
Since the nuclear spin-spin interactions between inequivalent protons result in a small
and random dipolar field, 
the fine structure of the $^1$H-NMR spectra produced by inequivalent protons was smeared out 
to form a rather broad spectrum of a two-peak structure, as shown in Fig. \ref{Fig1}. 
The spin-lattice relaxation rate $T_1^{-1}$ was measured at the center of the spectrum.

\begin{figure}[t]
\begin{center}
\includegraphics[trim=0.1cm 12.0cm 0.1cm 0.0cm, clip, width=80mm]
{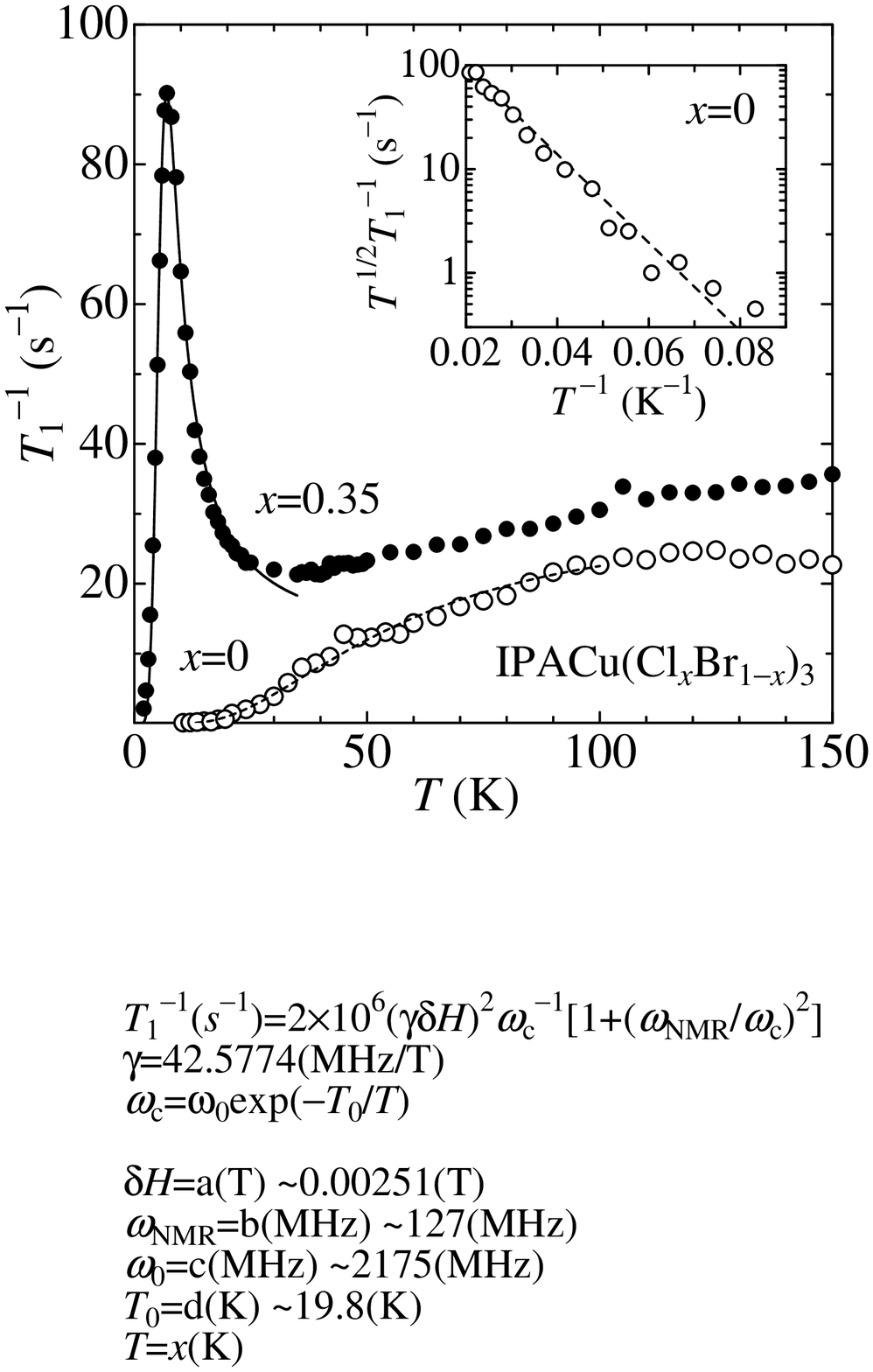}\\
%%{Fig1_4b.eps}\\
\end{center}
\caption{
Temperature dependence of spin-lattice relaxation rate $T_1^{-1}$ of 
IPACu(Cl$_x$Br$_{1-x}$)$_3$ with $x=0.35$ at $\nu=127$ MHz for $H \perp$ C-plane. 
The solid curve denotes the fit by eq. (\ref{eq3}). 
The temperature dependence of $T_1^{-1}$ for an $x=0$ system 
at $\nu=90.1$ MHz for the $H \perp$ B-plane is also shown for comparison. 
The dashed curve denotes the fit by eq. (\ref{eq2}),
where $\Delta$ was fixed at 98 K. 
The inset shows its Arrhenius plot. 
\label{Fig2}}
\end{figure}

First, we present the results of $^1$H-NMR spin-lattice relaxation rate $T_1^{-1}$ measurement
of the parent compound IPACuBr$_3$ measured at $\nu=90.1$ MHz 
for the $H \perp$ B-plane\cite{ManakaBr,Manakamixsus,Torque}. 
The nuclear magnetization recovery was fitted at all temperatures by the single exponential form
\begin{equation}
\label{eq1}
1- \frac{M ( \tau )}{M ( \infty )} \propto \exp \left( - \frac{\tau}{T_1} \right),
\end{equation}
where $M ( \tau )$, $M ( \infty )$, and $T_1$ are the nuclear magnetization at $\tau$, 
the thermal equilibrium value of the nuclear magnetization, 
and the spin-lattice relaxation time, respectively. 
The obtained temperature dependence of $T_1^{-1}$ is shown in Fig. \ref{Fig2}, 
where $T_1^{-1}$ shows a rapid decrease toward zero with decreasing temperature. 
We fitted the following equation below $T=100$ K:
\begin{equation}
\label{eq2}
T_1^{-1} \propto \frac{1}{\sqrt{T}} \exp \left( - \frac{\Delta}{k_{\rm B} T} \right),
\end{equation}
where $\Delta$ was fixed as 98 K estimated 
in the previous macroscopic experiments \cite{Manakamixsus}, 
as shown by the solid line in Fig. \ref{Fig2}. 
This equation indicates the existence of an excitation gap from the singlet ground state 
and was used for the estimation of the excitation gap $\Delta$ 
in the previous magnetic susceptibility and specific heat measurements \cite{Manakamixsus}. 
We can clearly see that the obtained temperature dependence of $T_1^{-1}$ in IPACuBr$_3$ 
was well reproduced by this equation, 
providing microscopic evidence of the gapped singlet ground state 
and a consistency with previous macroscopic experiments \cite{Manakamixsus}.

Next, we show the results of $^1$H-NMR spin-lattice relaxation rate $T_1^{-1}$ measurements 
in the $x=0.35$ system. 
Figure \ref{Fig1}(b) shows the typical nuclear magnetization recovery. 
As in the $x=0$ system, the nuclear magnetization recovery was 
well fitted at all temperatures by eq. (\ref{eq1}). 
Figure \ref{Fig2} shows that the $T_1^{-1}$ gradually decreases and then turns to 
increase below $T=30$ K with decreasing temperature. 
After that, the $T_1^{-1}$ peaks at $T=7.5$ K and then rapidly decreases toward zero. 
This behavior differs from that observed in the parent compound IPACuBr$_3$,
so we can conclude that the $T_1^{-1}$ peak structure is due to the bond-randomness effect 
in IPACu(Cl$_x$Br$_{1-x}$)$_3$ and that the ground state is 
different from the singlet-dimer phase of IPACuBr$_3$. 
This is in contrast to the conclusion drawn from macroscopic measurements\cite{Manakamixsus}. 
We also measured $T_1^{-1}$ for $\nu=63$ MHz in the $x=0.35$ system, 
and a similar behavior to that at $\nu=127$ MHz,
namely, that $T_1^{-1}$ showed a peak at $T=6.5$ K, 
was observed. 
The inset of Fig. \ref{Fig1} shows the temperature dependence of the resonance linewidth
in IPACu(Cl$_{0.35}$Br$_{0.65}$)$_3$. 
It shows no anomaly but only a slight and gradual increase between $T=$ 4 and 20 K. 
These results indicate that 
the observed peak structure of $T_1^{-1}$ is not due to the critical divergence 
indicative of the magnetic phase transition to the magnetic ordered phase, 
because the $^1$H-NMR spectra split with a width as large as 0.1 T 
when this system undergoes a long-range magnetic order at low temperatures
as observed in the $x =$0.15 system\cite{Kanada,KanadaQuBS}. 
A possibility of a magnetic order with an extremely small magnetic moment, 
that is, below 0.01 $\mu_{\rm B}$, which corresponds to the slight increase 
observed in resonance linewidth,  
is readily ruled out, because the existence of a static hyperfine field 
is denied by the muon experiment in this $x =$0.35 system\cite{Goto}.

The peak structure of $T_1^{-1}$ has been discussed theoretically 
by Bloembergen, Purcell, and Pound (BPP) \cite{BPP}. 
They showed that the divergence in $T_1^{-1}$ occurs when the NMR frequency is equal 
to the characteristic frequency of motions in the system. 
This means that the characteristic frequencies of magnetic fluctuation at 
$T=$ 7.5 and 6.6 K are 127 and 63 MHz, respectively, 
and hence that the magnetic fluctuation slows down with decreasing temperature. 
We have fit the observed peak with the BPP theory on the assumption
that the characteristic frequency $\omega_{\rm c}$ decays 
exponentially with decreasing temperature as $\omega_{\rm c}=\omega_{\rm c0}e^{-T/T_0}$,
where $\omega_{\rm c0}$ and $T_0$ are constants.
The fitted curve of the equation 
\begin{equation}
\label{eq3}
T_1^{-1} = \frac{2\cdot(\gamma \delta H)^2/\omega_{\rm c}}{1+(\omega_{\rm NMR}/\omega_{\rm c})^2},
\end{equation}
is shown in Fig. {\ref{Fig2}, where $\gamma$ is 
the proton nuclear gyromagnetic ratio (42.5779 MHz/T),
$\omega_{\rm NMR}$, the Lamor frequency, 
and $\delta H$, the amplitude of a fluctuating hyperfine field.  
The fit seems very good, and the
obtained parameters of $\delta H\simeq$ 25 Oe and $T_0 \simeq$ 20 K are quite reasonable.
In particular, the amplitude coincides with that obtained by the muon LF decoupling
measurement 38.5 Oe at 0.3 K\cite{Goto}. 
However, data taken at the two different NMR frequencies could not be fit with 
an identical $\omega_{\rm c0}$, indicating that 
the amplitude of the fluctuation is also temperature-dependent.
This fact excludes the possibility that the observed increase in $T_1^{-1}$ is 
due to the freezing of molecular rotation, because
when $T_1^{-1}$ is driven by slowing down of the rotation of molecules such as 
CH$_3$ or NH$_4$\cite{Endo}, its temperature dependence should be correctly
described by a single set of parameters.
We also ruled out the possibility of the spin-Peierls instability 
in one-dimensional systems, 
because in such a case, it has been confirmed both experimentally and theoretically that
$T_1$ becomes very short and temperature-independent owing to the phason fluctuation.
\cite{Peierls_NMR}

%%trim=left, bottom, right,up
%%h=here, t=top, b=bottom, p=separate figure page
%%clip, scale, width=0.9\linewidth

\begin{figure}[t]
\begin{center}
\includegraphics[trim=0.2cm 11.5cm 0.1cm 0.1cm, clip, width=80mm]
{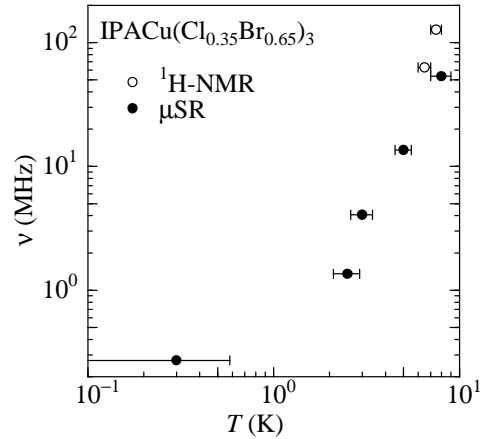}\\
%%{Fig2_2.eps}\\
\end{center}
\vspace*{-0.7cm}
\caption{
Temperature dependence of the characteristic frequency of magnetic fluctuation 
in IPACu(Cl$_{0.35}$Br$_{0.65}$)$_3$ estimated 
in the present $^1$H-NMR experiments 
and the recent muon spin relaxation experiments \cite{Goto}. \label{Fig3}}
\end{figure}
Quite recently, a muon spin relaxation experiment has also been carried out 
in the $x=0.35$ system, \cite{Goto} and it was observed 
through the longitudinal field (LF) decoupling of muon spin relaxation rate 
that the characteristic frequency of magnetic fluctuation decreases to zero,
indicating the critical slowing down, {\it i.e.}, the soft mode, 
toward the magnetic ground state, 
which is different from the singlet dimer phase concluded in previous macroscopic 
experiments \cite{Manakamixsus}. 
Figure \ref{Fig3} shows the temperature dependence of 
the characteristic frequency of magnetic fluctuation 
in the present system obtained from the present $^1$H-NMR experiments and 
the recent muon spin relaxation experiments \cite{Goto}. 

In the $^1$H-NMR experiments, the temperature at which the $T_1^{-1}$ has a peak 
for each frequency $\nu=63$ and 127 MHz is plotted. 
For muon spin relaxation experiments, the temperature where the muon spin relaxation rate has 
a maximum in the temperature dependence for each LF is plotted\cite{Goto}, 
because the muon spin relaxation is also driven by the fluctuating local field 
with a characteristic frequency the same as the Larmor frequency of muon spins 
$\nu = \gamma_{\mu} H_{\rm LF}$, where $\gamma_{\mu}$ and $H_{\rm LF}$ are 
the muon spin gyromagnetic ratio (135.534 MHz/T) and applied longitudinal field, respectively.
As shown in Fig. \ref{Fig3}, both temperature dependences 
of the characteristic frequency connect seamlessly, indicating that
the observed peak structure of $T_1^{-1}$ of $^1$H-NMR in the temperature dependence 
for the $x=0.35$ system also reflect the soft mode toward the absolute zero.

The appearance of a magnetic ground state without a long-range order
at finite temperatures, that is, an exotic ground state 
has been discussed by Fisher {\em et al.}\cite{Fisher}  
They investigated the disordered boson system and 
found that the ground state becomes the Bose-glass phase, which is characterized with 
the critical temperature of absolute zero and a massless and localized boson.  
Later, Oosawa {\em et al.} found that the ground state of the solid solution 
of two spin gap systems is a good candidate for Bose glass,
when one adopts mapping between a boson particle and vacuum in Fisher's notation and
the spin triplet and singlet sites in a spin system.  
Although both macroscopic and microscopic techniques have provided evidence of 
a Bose-glass phase in the (Tl,K)CuCl$_3$ 
quantum spin system\cite{OosawaTlK, Shindo, Suzuki, Fujiwara}, 
the reality of such a phase is still in question.
The soft mode observed in IPACu(Cl$_0.35$Br$_0.65$)$_3$ should give
strong evidence 
that the ground state of the present system is an exotic one 
represented by the Bose-glass phase.

Finally, we compare the present results in the $x=0.35$ system with the previous results 
in an $x=0.88$ system \cite{Adachi}. 
In the $x=0.88$ system, it was found that $T_1^{-1}$ has fast and slow components 
associated with gapless paramagnetic moment islands and gapped singlets, 
respectively, and that both $T_1^{-1}$s show no peak structure 
in their temperature dependences. 
We expect that this difference in $T_1^{-1}$ behavior between the present $x=0.35$ system 
and the $x=0.88$ 
system is related to the effect of nuclear spin diffusion on relaxation. 
In the $x=0.35$ system, which was found by muon experiments to consist of
magnetic islands and a singlet sea microscopically phase-separated \cite{Goto},
we infer that the mean distance of magnetic islands is too short 
that the nuclear spin polarization in an island is propagated by diffusion 
to neighboring islands before the completion of the spin-lattice relaxation, 
and hence, 
the nuclear magnetization recovery obeys the single exponential form as observed	
in the present experiments. 
This indicates that the magnetic islands in the $x=0.35$ system correlat more strongly
than those in the $x=0.88$ system, quite consistent with our observation
of the soft mode toward such a collective magnetic ground state in the $x=0.35$ system. 
Theoretically, in the ferromagnetic-antiferromagnetic random bond model investigated 
to explain the phase diagram of IPACu(Cl$_x$Br$_{1-x}$)$_3$, 
Nakamura \cite{Nakamura} found that there is a quantum critical point between the Haldane phase 
and the uniform antiferromagnetic ordered phase 
at the ferromagnetic bond concentration ratio $p=0.75$ 
corresponding to $x=0.87$ in the language of IPACu(Cl$_x$Br$_{1-x}$)$_3$. 
This is consistent with the experimental observation of the phase boundary at $x=0.87$. 
Nakamura also found that the ground state remains gapless 
including the uniform antiferromagnetic ordered phase for $0 < x < 0.87$, 
and even for $x \le 0.44$, which was characterized as the gapped singlet-dimer phase 
in previous macroscopic experiments \cite{Manakamixsus}. 
This indicates that the phase intrinsically differs between $x \le 0.44$ and $x \ge 0.87$, 
including the magnetic property, whether or not an energy gap exists. 
This difference may be related to our observation of the difference in $T_1^{-1}$ 
behavior between the present $x=0.35$ system and the previous $x=0.88$ system \cite{Adachi}. 

In summary, we have measured the spin-lattice relaxation rate $T_1^{-1}$ of $^1$H-NMR in 
(CH$_3$)$_2$CHNH$_3$Cu(Cl$_x$Br$_{1-x}$)$_3$ with $x=0$ and $0.35$, 
in order to microscopically investigate the magnetism of these systems, 
which have been reported to be of the singlet dimer phase from the previous macroscopic 
magnetization and specific heat measurements 
The temperature dependence of $T_1^{-1}$ in the $x=0$ system decreased exponentially 
toward zero, 
providing microscopic evidence of the gapped singlet ground state consistent 
with results of macroscopic experiments. 
At the same time, in the $x=0.35$ system, 
the temperature dependence of $T_1^{-1}$ showed a sharp peak structure
but no splitting of $^1$H-NMR spectra, 
indicative of the magnetic ordering observed at low temperatures. 
We conclude that this peak structure of NMR-$T_1^{-1}$ together with
the muon relaxation rate indicates
the soft mode toward the exotic magnetic ground state. 
Finally, we also note that the combination of these two microscopic probes 
enables a {\em wide-band detection}
of the spin fluctuation spectrum.

We thank T. Nakamura for helpful discussions. 
This work was supported by Grants-in-Aid for Scientific Research on Priority Areas 
``High Field Spin Science in 100 T'' 
from the Ministry of Education, Science, Sports and Culture of Japan, 
the Saneyoshi Scholarship Foundation, and 
the Kurata Memorial Hitachi Science and Technology Foundation.

\end{document}